\documentclass[a4paper,11pt]{article}
\pdfoutput=1 % if your are submitting a pdflatex (i.e. if you have
\usepackage{jheppub} % for details on the use of the package, please
\usepackage[utf8]{inputenc}
\usepackage{amsthm}
\usepackage{amssymb}
\usepackage{amsmath}

\usepackage{diagbox}
\usepackage{tikz}

\usepackage{feynmp}
\usepackage{tabularx}

\newcommand{\DEFT}{\texttt{DEFT}}
\newcommand{\abs}[1]{| #1 |}

\title{\DEFT: A program for operators in EFT}

\author[a]{Ben Gripaios}
\author[b]{and Dave Sutherland}

\affiliation[a]{University of Cambridge\\Cavendish Laboratory, J.~J.~Thomson Ave, Cambridge, CB3 0HE, UK}
\affiliation[b]{University of California Santa Barbara\\UCSB Broida Hall, Santa Barbara CA 93106-9530, USA}

\emailAdd{gripaios@hep.phy.cam.ac.uk}
\emailAdd{dwsuth@ucsb.edu}

\abstract{We describe a \texttt{Python}-based computer program, \DEFT,  for
  manipulating
  operators in effective
  field theories (EFTs). In its current incarnation, \DEFT\ can be
  applied to 4-dimensional, Poincar\'{e} invariant theories with gauge group
  $SU(3)\times SU(2) \times U(1)$, such as the Standard Model (SM), but a
  variety  of extensions (e.g. to lower dimensions or to an arbitrary product of
unitary gauge groups) are conceptually straightforward.
Amongst other
  features, the program is
  able to: (i) check whether an input list of Lagrangian operators (of a given
  dimension in the EFT expansion) is a
  basis for the space of operators contributing to S-matrix elements, 
 once redundancies
  (such as Fierz-Pauli identities, integration by parts, and equations of
  motion) are taken into account; (ii) generate such a basis (where
  possible) from an input algorithm; (iii) carry out a change of
  basis. We describe applications to the SM (where we
  carry out a number of non-trivial cross-checks) and extensions thereof, and outline how the program may be
  of use in precision tests of the SM and in the ongoing search for new physics at the LHC
  and elsewhere. The code and instructions can be downloaded from
  \url{http://web.physics.ucsb.edu/~dwsuth/DEFT/}.}

\newcommand{\op}{\mathcal{O}}
\newcommand{\lag}{\mathcal{L}}

\newcommand{\rr}{\mathbb{R}}
\newcommand{\nn}{\mathbb{N}}

\newcommand{\al}{\alpha}
\newcommand{\be}{\beta}
\newcommand{\ga}{\gamma}
\newcommand{\del}{\delta}
\newcommand{\eps}{\epsilon}

\begin{document} 
\maketitle
\flushbottom
\section{Introduction}

Non-renormalizable
quantum field theories, once regarded as something of a pariah by physicists, have
become ubiquitous as a means of parameterizing, in a general way, the
low-energy effects of
unknown physics residing at higher-energy scales. In a nutshell, given a set of quantum fields representing physical
degrees of freedom and a group of symmetries acting on them, the
lagrangian of such an effective field
theory (EFT) contains not just
renormalizable invariant operators built out of fields and spacetime
derivatives, but all invariant operators, ordered by their relevance in terms of a low-energy expansion.
The example that is perhaps of greatest current interest to particle physicists is the use of
an EFT given by the renormalizable Standard Model (SM) plus higher-dimension
operators (henceforth, the `SMEFT'), to parameterize possible deviations from
the SM at the Large Hadron
Collider and elsewhere. 

Whilst the idea of EFT is simple enough in principle, to use it in practice
involves a great deal of donkey work, above and beyond what is required
in renormalizable quantum field theory. There are several reasons for this. Firstly, the operators at a given order in the
low-energy expansion form a vector space whose dimension grows
exponentially with the order. Secondly, given some set of physical
observables (which, as we shall see, may be considered to span a
subspace of the dual vector space), there is a large subspace of
operators (whose dimension also grows
exponentially with the order of expansion) that are redundant, in the sense that they do not
contribute to any of the observables. These operators must be
identified and dealt with, by forming a quotient space of physical operators \cite{Einhorn:2013kja}. Thirdly, in fitting either to
experimental data or to some overarching theory, one
must choose a basis for the space of physical operators. Different data sets
and different theories prefer different bases (as do different
physicists!) and comparison between
them necessitates a change of basis.

As we shall explain in more detail in \S \ref{sec:pre}, much of the required
donkey work reduces to combinatorics and linear algebra, and is
easily done with a computer. To this end, in this work we present a
computer code, \DEFT, to help with the work.

In its current implementation,
\DEFT\ can be applied `out-of-the-box' only to the SMEFT, but the methods employed are
easily generalized to EFTs with arbitrary field content, in which the
symmetry group is an arbitrary product of unitary groups. The code
could thus easily be generalized to apply to a number of other EFTs
of potential physical interest. To give one obvious example, given anomalies in data suggesting the
need to add light, beyond-the-SM degrees of freedom to the SM, it
would be a
simple matter to incorporate such fields into \DEFT. This has already been
done \cite{Gripaios:2016xuo} in the case of a SM gauge singlet scalar, hypothesized to
explain a spurious anomaly in the $\gamma \gamma$ spectrum at an
invariant mass around 750 GeV. To give another example, by use of a sigma-model field transforming as a
bi-fundamental under $SU(n)\times SU(n)$, \DEFT\ could easily be
adapted to apply to the
chiral lagrangian describing QCD with $n$
light quarks at hadronic energy scales. Similarly,  \DEFT\ could
be adapted to use in flavour physics, where the relevant effective lagrangian at
the scale of $b$-quarks has $SU(3) \times U(1)$ invariance (where
$U(1)$ corresponds to the electromagnetic gauge group) with the Higgs boson,
$W,Z$ bosons, and top quark removed. \DEFT\ could also be used to
  evaluate the restricted set of invariants that arise in theories
  with some unified symmetry group, such as the $SU(4)\times SU(2)
  \times SU(2)$ of Pati \& Salam or the $SU(5)$ of Georgi \& Glashow. With a bit more effort, \DEFT\ could also be applied to theories in lower dimensions, or
indeed those with Galileo, rather than Poincar\'{e}, invariance.

In rough terms, \DEFT\ does the following. At each given order in
the EFT dimension expansion, \DEFT\ generates all possible
lagrangian invariants. As already indicated, these may be considered
to form a vector space, $V$,
over the real numbers. Given a space of observables (which may be
regarded as linear maps from $V$ to the reals, and hence as elements
of the dual space of $V$) one may define a
subspace of redundant operators, $W \subseteq V$, as those that do not affect
measurements of the observables. The quotient space $U \equiv V - W$
represents the space of
physical operators. \DEFT\ generates the subspace of redundant
operators for the case in which the space of observables consists of
the whole
$S$-matrix. Given a set of vectors in $V$ (perhaps defined by some
algorithm based on the user's preference), \DEFT\ will check
whether the equivalence classes in $U$ containing those vectors are
linearly-independent and span
$U$, and hence may be used to form a basis of physical
operators. Given two such bases, \DEFT\ will provide an explicit
formula for the change of basis  in $U$. 

Experienced practitioners of EFT will easily be able to imagine the benefits of
an automated approach of this type, but let us spell a few of them
them out anyway. 

Firstly, \DEFT\ is able to generate a basis of
operators at a given dimension that is not only (hopefully) correct,
but is also obtained relatively quickly, provided that the
  operator dimension
is not too large. For example, for
the SMEFT with one generation of fermions at operator dimension six,
\DEFT\ generates the list of 84 operators in Fig.~\ref{fig:crappamundi} in a matter of minutes. This is to be contrasted with the human approach, which
took roughly a quarter of a century, with more than one hiccough along
the way \cite{Buchmuller:1985jz,Grzadkowski:2010es}.

Secondly, there is a large freedom in the choice of operator
  basis, which \DEFT\ enables the user to exploit, according to his or
  her particular {\em desiderata}. There are two aspects to this
  freedom. The first corresponds to the usual freedom to choose a
basis for a vector space. But in EFT, there is yet more freedom, which corresponds to the
fact that the underlying physical object is a quotient vector space. A
vector in the quotient space, $U$, (in particular a basis
vector), can be represented by any vector in $V$ that lies in the corresponding
equivalence class. It is often useful, in applications, to exploit
this freedom. On the one hand, for example, an experimentalist whose apparatus is only
able to detect certain types of particles, might prefer a basis
description which prioritises operators containing those
particles. On the other hand, an experimentalist whose apparatus detects only
very low energy particles might prefer a basis description with
operators containing as few derivatives as possible. Given some
input algorithm encoding the user's {\em desiderata}, \DEFT\ will
output a corresponding basis (and check that it is indeed a basis, in
the sense that it is a linearly-independent spanning set for $U$). So
for example, in the SMEFT at dimension 6, the user could simply input
a list of 84 operators, and ask \DEFT\ to check that it is a
basis. Or the user could input his favourite 10 operators
and ask \DEFT\ to generate (if possible) 74 others using its default
algorithm or some modification thereof.\footnote{The README file gives some indication as to how bases may be input.}

This freedom to choose a basis has its downsides, of course. 
Indeed, it seems to be an empirical law of nature that,
given an EFT that describes the low-energy limits of some
theorists' models and which is subject to the constraints of some
experimentalists' measurements, the relevant literature is likely to
contain roughly as many different choices of basis as the number of
theorists and experimentalists put together! This is hardly
surprising: the former are likely to choose bases in which the
particular operators their theories generate are basis elements and
the latter are likely to choose bases in which the operators they
constrain best are basis elements. For a few examples of the
proliferation of such bases in the SMEFT at $d=6$, the reader is
invited to consult, e.g.~\cite{Falkowski:2015wza}. As a result, the community has
arrived at something of an impasse: in order to compare
theory with experiment, or indeed to just compare one experiment with
another, phenomenologists must be able to change bases.
But such changes of bases are highly non-trivial, because the aforementioned redundancies among
operators must be taken into account. To be explicit, one wishes to carry out changes of basis in
the quotient space $U$ of physical operators, but using a description in
terms of operators in $V$, resulting in a computation that is rather
more computationally intensive than might be suggested by the
dimensions of either of $U$ or $V$ alone. Indeed, thus far just one such change of basis has
been carried out (by hand) in the SMEFT at dimension 6
\cite{Falkowski:2001958}.

A third, and perhaps the most significant, benefit of \DEFT\ is that
such changes of basis can be carried out, not quite at the
touch of a button, but with comparable ease. As an example, we
describe the use of \DEFT\ to carry out a change of basis in the SMEFT
at dimension 6 in \S\ref{sec:cob}. The computation takes 20 minutes on a laptop.
We hope therefore, that in removing this impasse \DEFT\ will prove to
be useful in
the current programme of comparing experimental data with the SM via EFT.

The ability of \DEFT\ to construct arbitrary bases and change between them gives it
something of an
advantage with respect to recent analytic efforts to determine an EFT basis
using Hilbert series methods \cite{Lehman:2015via,Henning:2015daa,Lehman:2015coa,Henning:2015alf,Henning:2017fpj}. While these methods are
extremely elegant, they naturally require a specific type of basis, namely one
in which the numbers of derivatives appearing in operators are
minimized. \DEFT\ also enables us to perform an
independent cross-check of these methods.

The genericity of \DEFT\ also distinguishes it from existing \texttt{Python} frameworks with practical applications to (SM)EFT, into which are encoded mappings between particular bases \cite{Falkowski:2015wza}, or explicit transformation rules for the conversion between different equivalent operators \cite{Criado:2017khh}. For this reason we envisage one use of \DEFT\ to be the construction or conversion between bespoke bases in the Standard Model or similar field theories.

The main drawback of \DEFT\ is that it rapidly runs out of steam as
the operator dimension grows. This is hardly surprising, since \DEFT\
works by performing brute-force linear-algebra
manipulations in vector spaces whose dimension grows exponentially
with the operator dimension, in terms of a redundant description whose
size also grows exponentially. Given current computing capabilities,
the ceiling corresponds to spaces with roughly $10^3$ physical
operators. So in the one-generation SMEFT, going beyond dimension 9 is inconceivable.

The outline of the rest of the paper is as follows. In the next Section, we develop the
required mathematical formalism in more detail. In \S \ref{sec:nuts} we describe the implementation in the code and its use, and in \S \ref{sec:cross} we
describe a number of examples and cross-checks.
%%%%%%%%%%%%%%%%%%%%%%%%%%%%%%%%%%%%%%%%%%%%%%%%%%
\section{Mathematical preliminaries \label{sec:pre}}
%%%%%%%%%%%%%%%%%%%%%%%%%%%%%%%%%%%%%%%%%%%%%%%%%%
Though ultimately all of the computations \DEFT\  carries out
will be done in a
specific basis or bases, we find it helpful to begin by framing the
discussion in a way that is basis independent.
We thus define the {\em operators} of a given dimension as the gauge
and Poincar\'{e} invariants built out of formal combinations of fields and spacetime
derivatives. We insist that these be hermitian.\footnote{We remark that it
  is common in the literature to allow non-hermitian operators (with
  correspondingly complex coefficients) such that the resulting
  lagrangian is hermitian. Since this confuses the counting of operators,
  we insist that they be hermitian for counting purposes.} Since the sum of two
such operators is itself an operator, and since multiplying an
operator by a real number yields an operator, {\em \&c}, the operators
form a (finite-dimensional) vector
space, $V$ over the reals.

At a given order in the expansion, the {\em observables} may be
regarded as
maps from $V$ to $\mathbb{R}$, where the value of the map is given by the
real number that would be obtained by a measurement of the observable, given the theory corresponding to that operator. At the given order,
the 
operator contributes to the observable via interference with lower
dimension operators, and so the map is linear. Hence
the observables are elements of the dual vector space $V^*$.

Not every element of $V^*$ (not every linear map) can be an observable, however. For one
thing, the operators in $V$, which we regard as formal combinations of fields
and derivatives, may be subject to underlying mathematical
identities. For example, some linear combination of operators may satisfy a Fierz or
Schouten identity, or be a total derivative. As a result (at least in
perturbation theory), all observables must yield zero on those linear
combinations of formal operators. 
Moreover, at least in particle physics
collider experiments
(although not necessarily in other areas of physics), observables are
restricted to $S$-matrix observables (things which can be measured `at
infinity') and one may show ({\em cf.} \S \ref{sec:eq}) that such observables yield zero for any operator that vanishes when the renormalizable equations of motion
hold, up to corrections that are higher order in momentum counting. Finally, it may happen that, given our current
technological limitations, some things are simply not observable, or
that we are simply not interested in them.
Thus it is useful to define a subspace $U^* \subset V^*$ of
observables of interest. 

Given $U^*$, it is natural to consider the space $W \subset V$ of {\em
  redundant operators} defined as the
operators that yield zero for all observables. It then follows that $U
\cong (U^*)^*$ is simply given by $V-W$, the quotient space obtained
by identifying any two operators in $V$ that differ by an operator in $W$. $U$ is also a vector
space (though it is not a subspace of $V$!) and we call it the {\em
  space of physical operators}. We stress (as in \cite{Einhorn:2013kja}) that the
elements of $U$ are equivalence classes of operators in $V$, where the
equivalence relation is defined such that any two operators in $V$
whose difference lies in $W$ are considered equivalent.  

We stress again that,
according to our definition, $U^*$ includes not only observables that are
`mathematically unobservable', in the sense of being related by underlying
identities that hold irrespective of what we do and do not observe, but also those that are unobservable because of the
restricted nature
of the experiments that we have in mind. We find this to be a useful
concept, as the following examples illustrate. 

Suppose, for example, that, much like the ancient Greeks, our
experiments are purely of the {\em gedanken} variety, such that we
don't bother to 
measure anything. Then $U^* = \{0\}$ contains only the zero vector, $W = V$, and $U=\{0\}$, such that there
are no non-trivial physical operators. 

Alternatively, suppose that we are only interested in searching for baryon number
violation by $\pm$ 2 units at dimension 6 in the SMEFT, and so restrict our attention to experiments
sensitive to processes in which baryon number is violated by $\pm$ 2 units. Then $W$
contains all operators in which baryon number is violated by some
other number of units, because
at this order, such operators can only interfere with lower dimension
operators in the SMEFT, all of which conserve baryon number and so
lead to a violation of baryon number by a number of units which is not
equal to $\pm$ 2. The
physical operators in $U$ are then those operators which violate
baryon number by $\pm$ 2 units.\footnote{This example admits the
  following generalization. At a given dimension, we may consider a symmetry of the lagrangian at lower dimensions, accidental or otherwise, and reduce operators into
combinations carrying
real  irreducible representations of that symmetry. (The representations are real because the operators in
  the lagrangian are
  elements of a real vector space. Hence the need to consider
  processes violating baryon number by either +2 or -2 units in the
  example.) Any collection of
these irreps can be associated with a corresponding subspace of
observables in a similar way.}

Finally, suppose that we build a `superdupercollider' in which all
$S$-matrix elements are observable. The corresponding $U$, which we
  will consider exclusively henceforth, contains all
operators that do not vanish under mathematical identities or when the
equations of motion hold.  

As described in the introduction, our main goals are to find automatic
procedures for generating and characterising the spaces $V$, $W$, and
$U$. In particular, we would like to be able to find explicit, {\em
  bona fide} bases
for $U$, which can then be used to fit data to experiment, and to be
able to perform a change of basis, such that fits performed using
different bases can be compared.

Before we do so, we make a few parenthetic remarks on truncations of
the space of operators, which are often carried out in the literature.
%%%%%%%%%%%%%%%%%%%%%%%%%%%%%%%%%%%%%%%%%%%%%%%%%%
\subsection{Remarks on truncations \label{sec:trunki}}
%%%%%%%%%%%%%%%%%%%%%%%%%%%%%%%%%%%%%%%%%%%%%%%%%%
Since the dimension of the space of physical operators grows
exponentially with the order of the EFT expansion, it tends to quickly
become unmanageably large. For example, in the SMEFT with 1 generation
of fermions, the space is 1 dimensional in $d=0$ and $2$
(corresponding to the vacuum energy and Higgs mass parameter,
respectively, and 19,\footnote{The number is reduced to 17 if one
  eliminates the operators $B \tilde B$ and $W \tilde W$; these
  are retained in {\tt DEFT}.} 84, and 993 dimensional in $d=4,6,8$
respectively \cite{Grzadkowski:2010es,Henning:2015alf}. 

Given this state of affairs, it is natural to try to reduce the
dimension of the space by some kind of truncation. We have already
shown how this can be done by restricting to the space of physical
observables of interest and defining a corresponding space of
redundant operators as those which do not contribute to the
observables of interest. The space of physical operators is then
obtained as the quotient space.

Many authors have gone further, in restricting to a subspace of
operators on the basis of some kind of theoretical prejudice. Though
it is somewhat out of the main thrust of this paper, we feel it
worthwhile to issue some parenthetical remarks regarding the pitfalls
of such an approach. 

To be concrete, the typical strategy is to pick a
`theoretically-motivated' list of operators and then to consider just the
span of those operators in $V$ in fits to data. Now, it is certainly
the case that such a span defines a subspace of $V$ and,
correspondingly, a subspace of the space of physical operators
$U$. Each vector in the latter subspace represents a perfectly viable
theory (at least from the EFT perspective) and so one may sensibly ask
whether the data rule it out or not. But one should be very careful in
trying to assign some physical meaning to the span of the operators as
a whole. Indeed, such a meaning can only be unambiguous if it is
well-defined on $U$, {\em i.e.} on the equivalence classes in $V$. 

It is perhaps easiest to illustrate the danger by means of an explicit
example. Suppose, for example, (as has been done in the literature)
that one is interested in the possibility of new physics effects in
the top quark sector. Given a basis of operators for $V$, one could then try
to truncate by retaining only operators featuring a top quark field in
that basis. But such a truncation certainly does not correspond to the
class of theories with new physics in the top sector, because it is
not well-defined on the equivalence classes of physical operators!
Indeed, it is a choice which depends arbitrarily on the basis that we
choose for $V$. If we change to a basis in which we replace an operator
involving a top quark with an operator not involving a top quark, then
the truncated space of physical operators that we obtain will also
change.

For another example, suppose that we try to divide operators into the
order at which they can be generated in a renormalizable UV
completion. So, for example, we might consider only the operators that
can be generated at tree-level. But the meaning of this is
ambiguous, because it is not, in general, well-defined on the
equivalence classes. Indeed, a number of counterexamples in the SMEFT
are given in \cite{Einhorn:2013kja}.

These ambiguities can be avoided by truncating directly on
the equivalence classes themselves. The problem, of course, is that
the equivalence classes are rather difficult to characterise. 
\DEFT\ can be used to help with the characterization. As an
  example of this, in \S\ref{sec:cross} we provide a spanning set of unobservable directions in the SMEFT and describe some of their properties.

We have already given one example of a manifestly consistent
truncation, namely in dividing operators into the real, irreducible
representations they carry of the symmetries (accidental or otherwise)
of the lower-order lagrangian. Thus, in the SMEFT at $d=5$ we may
classify the operators according to the representations of baryon and
lepton family numbers that they carry, while at $d=6$ we may classify
them by their baryon number and lepton parity; for $d\geq 7$ no accidental
symmetries remain.

Many other possible truncations remain. Indeed, any truncation of the space
of physical operators will do. But, presumably, some of these
truncations are more natural
than others. For example, an inspection of
the redundancy relations shows that the $d=6$ operator containing
 fields $GGHH$ is in
a class of its own, such that it always appears in a basis. We provide
an argument for this based on the general structure of operator redundancies described in Appendix \ref{app:relstruct}.
%%%%%%%%%%%%%%%%%%%%%%%%%%%%%%%%%%%%%%%%%%%%%%%%%%
\section{Implementation \label{sec:nuts}}
%%%%%%%%%%%%%%%%%%%%%%%%%%%%%%%%%%%%%%%%%%%%%%%%%%

From a (possibly overcomplete) ordered list of hermitian operators $\{\op_j\}$, we construct the most general lagrangian term
\begin{equation}
\op = \sum_j c_j \op_j,
\end{equation}
where $c_j \in \rr$ are real (Wilson) coefficients which define coordinates for the vector $\op$, which is an element of $V$. Without loss of generality, any observable linear in these coordinates is a real number, written as
\begin{equation}
\text{obs.} = \sum_j \alpha^j c_j,
\end{equation}
for some real $\alpha^j$, which define an element of $U^*$. If the original list of operators is overcomplete,
there exist directions $r^n_j$ --- elements of $W$ --- which satisfy $\sum_j \alpha^j r^n_j =
0$ for any observable quantity. We construct a matrix $M_{ij} =
r^i_j$ where each row is an unobservable direction in the original
list of operators; the rank of $M$ determines the dimension of the space $W$, i.e.~how many operators we may eliminate from the original list to form a basis.

To determine a spanning set of class representatives in $U$, we put the matrix $M$ in
reduced row echelon form (RREF).\footnote{A matrix is in reduced row echelon
  form iff. the leading coefficient (the first non-entry from the
  left) in each row is a $1$, each such leading coefficient is the
  only non-zero entry in its column, and each leading coefficient is
  to the right of that of the row above it. Its form is invariant
  under row operations on the original $M_{ij}$.} For each row in the
RREF matrix, we remove the operator corresponding to the column of the row's leading
coefficient --- whichever operators then remain are the equivalence class representatives of a viable basis, a.k.a the basis operators. Conveniently, each row also yields an expression for the
removed operator in terms of the basis operators.

As a concrete example, consider an ordered list of five operators $\{\op_i\}$, $i=1,\ldots,5$, and four unobservable directions $r^n_j$ between them, of which three are independent, leaving a two-dimensional basis. The RREF of $M$ is schematically
\begin{equation}
M_{ij} = \begin{pmatrix}
* & * & * & * & * \\
* & * & * & * & * \\
* & * & * & * & * \\
* & * & * & * & * \end{pmatrix}
\overset{\text{RREF}}{\rightarrow} \begin{pmatrix}
1 & r_{11} & 0 & 0 & r_{21} \\
0 & 0 & 1 & 0 & r_{22} \\
0 & 0 & 0 & 1 & r_{23} \\
0 & 0 & 0 & 0 & 0 \end{pmatrix}
\label{eq:rrefexample}
\end{equation}
such that we may choose $\op_2$ and $\op_5$ to be the class representatives of the basis. To express any lagrangian $\lag = \sum_{i=1}^5 c_i \op_i$ in terms of the basis operators $\lag = b_a \op_2 + b_b \op_5$, we may rearrange the non trivial elements of the RREF matrix, yielding
\begin{equation}
\begin{pmatrix} b_a \\ b_b
\end{pmatrix}
= R_{\alpha j} c_j =
\begin{pmatrix} 
-r_{11} & 0 & 0 & 1 & 0 \\
-r_{21} & -r_{22} & -r_{23} & 0 & 1
\end{pmatrix}
\begin{pmatrix}
c_1 \\ c_3 \\ c_4 \\ c_2 \\ c_5
\end{pmatrix} \, ,
\end{equation}
for some matrix $R$.

Suppose we choose another pair of class representatives, $\op^\prime_a$ and $\op^\prime_b$, which can be expressed in terms of the original monomial operators as $\op^\prime_\alpha = S_{\alpha j} \op_j$. Then we can change bases, i.e., relate the coefficients of the lagrangian $\lag = b_a^\prime \op^\prime_a + b_b^\prime \op^\prime_b$ to their unprimed counterparts, via
\begin{equation}
b_\alpha = (R S^T)_{\alpha \beta} b_\beta^\prime .
\end{equation}
The $2 \times 2$ matrix $R S^T$ is invertible iff. the primed operators form a complete basis, and one may thereby convert between arbitrary bases via conversion to and from the original unprimed basis.

We now describe how to construct a suitable matrix $M$, i.e., how to generate an overcomplete list of operators and redundant relations between them.

\subsection{Constructing operators}

\DEFT\ assumes that fields transform in irreps of $SU(N)$, which are
described via a combination of upper and lower indices with symmetry
conditions attached. An upper index takes values between $1$ and $N$
and transforms in the defining rep of $SU(N)$; a lower index runs
between $1$ and $N$ and transforms in the conjugate of the defining
rep. Conjugation of a field in an irrep of $SU(N)$ lowers upper
indices and vice versa. Presently, \DEFT\ contains the definitions for the fundamental and anti-fundamental irreps, along with the symmetric and traceless combinations thereof.

For our purposes, the irreps of the Lorentz group are those of $SU(2)_\text{L,lor} \times SU(2)_\text{R,lor}$ --- represented by the familiar undotted and dotted indices for the respective $SU(2)$s of the direct product --- with the distinction that, upon conjugation of a field, undotted indices are dotted and vice versa. Figure \ref{fig:smfieldreps} contains the Lorentz and gauge representations, as well as their explicit realisations in terms of (anti)fundamental indices, of the fields of the one generation Standard Model.

\begin{figure}
\begin{tabular}{c | c | c | c | c | c | c |}
Field & Dimension & $SU(3)_c$ & $SU(2)_L$ & $U(1)_Y$ & $SU(2)_\text{lor,L}$ & $SU(2)_\text{lor,R}$ \\ \hline
${L_L}^{\al a}$ & $\frac{3}{2}$ & $1$ & $2$ & $-\frac{1}{2}$ & $2$ & $1$ \\
${e_R}^{\dot\al}$ & $\frac{3}{2}$ & $1$ & $1$ & $-1$ & $1$ & $2$ \\
${Q_L}^{\al a A}$ & $\frac{3}{2}$ & $3$ & $2$ & $\frac{1}{6}$ & $2$ & $1$ \\
${u_R}^{\dot\al A}$ & $\frac{3}{2}$ & $3$ & $1$ & $\frac{1}{3}$ & $1$ & $2$ \\
${d_R}^{\dot\al A}$ & $\frac{3}{2}$ & $3$ & $1$ & $-\frac{2}{3}$ & $1$ & $2$ \\
$H^{a}$ & $1$ & $1$ & $2$ & $\frac{1}{2}$ & $1$ & $1$ \\
$B_{(\al \be)}$ & $2$ & $1$ & $1$ & $0$ & $3$ & $1$ \\
$W^a_{b (\al \be)}$, $W^a_{a (\al \be)} \equiv 0$ & $2$ & $1$ & $3$ & $0$ & $3$ & $1$ \\
$G^A_{B (\al \be)}$, $G^A_{A (\al \be)} \equiv 0$ & $2$ & $8$ & $1$ & $0$ & $3$ & $1$ \\
\end{tabular}
\caption{The fields of the one generation Standard Model in component form, along with their mass dimensions, and their representations under the SM symmetries.\label{fig:smfieldreps}}
\end{figure}

The advantage of working with exclusively fundamental and anti-fundamental indices is that there are only two invariant tensors: the Kronecker delta $\delta^a_b$ (with an upper and lower index) and the Levi-Civita epsilon $\eps^{a b c \ldots z}$ or $\eps_{a b c \ldots z}$ (with either $N$ upper or $N$ lower indices). We report various sign conventions in Appendix \ref{app:conv}.

To form all monomial singlet operators from an input set of fields,
\DEFT\ generates all combinations of fields, their conjugates, and covariant derivatives acting thereon\footnote{A covariant derivative $D_{\al \dot \al}$ has a lower $SU(2)_\text{L,lor}$ and a lower $SU(2)_\text{R,lor}$ index.} satisfying a specified boolean method (usually that the mass dimension of the putative operator is less than or equal to a given number). The list of operators is filtered for zero net $U(1)$ charges. The $SU(N)$-like indices of each operator are then partitioned by group and contracted with all combinations of deltas and epsilons. These contractions are filtered according to whether they respect the symmetry conditions of the fields' indices (e.g.~an epsilon tensor cannot contract two indices which are symmetrized, and a delta tensor cannot cannot an upper and lower index which are required to be traceless).

Having generated a (over)complete set of monomial operators, we must now generate a (over)complete set of linear combinations of the monomial operators which do not contribute to scattering amplitudes. Those for the Standard Model fall into the four categories of \S\S\ref{sec:ibp}-\ref{sec:eq}.

\subsection{Integration by parts\label{sec:ibp}}
Amplitudes which are proportional to a sum of the momenta of the external legs --- $\sum_{i \in \text{external}} k^\mu_i M_\mu$ for external leg momenta $\{k^\mu_i\}$ --- are zero by overall momentum conservation. At the operator level, for each term $D_{\al \dot \al} F D_{\be \dot \be} G \ldots D_{\ga \dot \ga} D_{\del \dot \del} H$, we generate a relation by moving the outermost derivative of each field, i.e.,
\begin{align}
D_{\al \dot \al} (F D_{\be \dot \be} G \ldots D_{\ga \dot \ga} D_{\del \dot \del} H) &= D_{\be \dot \be} (D_{\al \dot \al} F  G \ldots D_{\ga \dot \ga} D_{\del \dot \del} H) = \ldots \nonumber \\
&= D_{\ga \dot \ga} (D_{\al \dot \al} F D_{\be \dot \be} G \ldots D_{\del \dot \del} H) = 0 \, .
\end{align}
\subsection{Fierz relations\label{sec:fierz}}
A product of one upper and one lower Levi-Civita epsilon tensor may be expressed as a sum of products of Kronecker deltas:
\begin{equation}
  \epsilon_{ab \cdots c} \eps^{x y \cdots z} + \sum_{\xi \in S_N} \sigma(\xi) \delta_a^{\xi(x)}\delta_b^{\xi(y)} \ldots \delta_c^{\xi(z)} = 0,
\label{eq:fierz}
\end{equation}
summing over the permutations $\xi$ of the $N$ upper indices, each having parity $\sigma(\xi) = \pm 1$. For $N=2$, there are also Schouten identities,
\begin{align}
 \eps_{ab} \eps_{cd} - \eps_{ac} \eps_{bd} + \eps_{ad} \eps_{bc} &= 0 ,\\
 \eps^{ab} \eps^{cd} - \eps^{ac} \eps^{bd} + \eps^{ad} \eps^{bc} &= 0 ,\\
 \del^a_b \eps_{cd} - \del^a_c \eps_{bd} + \del^a_d \eps_{bc} &= 0 ,\\
 \del_a^b \eps^{cd} - \del_a^c \eps^{bd} + \del_a^d \eps^{bc} &= 0 ,
\end{align}
which are effectively `raised' and `lowered' versions of
(\ref{eq:fierz}). \DEFT\ searches for the lefthandmost term in each operator, and generates one relation for each match.

In addition, any $SU(N+k), k \in \nn$ relation of the form (\ref{eq:fierz}) may have its indices restricted to run between $1$ and $N$, yielding an $SU(N)$ Fierz relation
\begin{equation}
  \sum_{\xi \in S_{N+k}} \sigma(\xi) \delta_a^{\xi(x)}\delta_b^{\xi(y)} \ldots \delta_c^{\xi(z)} = 0.
\end{equation}
For each set of operators with the same field content having $N_u$ upper and $N_l$ lower $SU(N)$ indices, one such relation is generated for each $k \leq \min(N_u,N_l) - N$.
\subsection{Commuting covariant derivatives\label{sec:comm}}
For a field $G^{ab\cdots}_{xy\cdots}$ which transforms under an $SU(N)$ gauge group with upper indices $a,b,\ldots$ and lower indices $x,y,\ldots$, one can interchange any two of its adjacent covariant derivatives, and the difference of the terms is a sum of field strengths:
\begin{align}
  (D \ldots D) [ D_{\al \dot \al},D_{\be \dot \be} ] (D \ldots D) G^{ab\cdots}_{xy\cdots} = i g_N (& {F_{\al \dot\al \be \dot\be}}^a_t G^{tb\cdots}_{xy\cdots} + {F_{\al \dot\al \be \dot\be}}^b_t G^{at\cdots}_{xy\cdots} + \ldots \nonumber\\
 - & {F_{\al \dot\al \be \dot\be}}^t_x G^{ab\cdots}_{ty\cdots} - {F_{\al \dot\al \be \dot\be}}^t_y G^{ab\cdots}_{xt\cdots} - \ldots),
\end{align}
where $g_N$ and ${F_{\al \dot\al \be \dot\be}}^a_t = -\frac{1}{2}  \eps_{\dot\al \dot\be} {F_{\al \be}}^a_t - \frac{1}{2} \eps_{\al\be} {\overline{F}_{\dot\al \dot\be}}^a_t$ are respectively the gauge coupling and field strengths of the $SU(N)$ gauge group. One relation is generated per operator per field per adjacent pair of covariant derivatives.
\subsection{Equations of motion\label{sec:eq}}
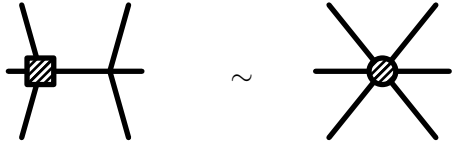
\begin{figure}
\begin{center}
\begin{tabular}{m{25mm} m{7mm} m{25mm}}
\begin{fmffile}{eomgraph}
\begin{fmfgraph*}(50,50)
\fmfpen{thick}
\fmfleftn{l}{3}
\fmfrightn{r}{3}
\fmf{plain}{l1,v1}
\fmf{plain}{l2,v1}
\fmf{plain}{l3,v1}
\fmf{plain,tag=1}{v1,v2}
\fmf{plain}{v2,r1}
\fmf{plain}{v2,r2}
\fmf{plain}{v2,r3}
\fmfv{d.sh=square,d.filled=shaded,d.size=.2w}{v1}
%\fmfposition
%\fmfipath{p[]}
%\fmfiset{p1}{vpath1(__v1,__v2)}
%\fmfiv{d.sh=cross,d.ang=0}{point length(p1)/3 of p1}
\end{fmfgraph*}
\end{fmffile}
& $\sim$ &
\begin{fmffile}{eomgraph2}
\begin{fmfgraph*}(50,50)
\fmfpen{thick}
\fmfleftn{l}{3}
\fmfrightn{r}{3}
\fmf{plain}{l1,v1}
\fmf{plain}{l2,v1}
\fmf{plain}{l3,v1}
%\fmf{plain,tag=1}{v1,v2}
\fmf{plain}{v1,r1}
\fmf{plain}{v1,r2}
\fmf{plain}{v1,r3}
\fmfv{d.sh=circle,d.filled=shaded,d.size=.2w}{v1}
%\fmfposition
%\fmfipath{p[]}
%\fmfiset{p1}{vpath1(__v1,__v2)}
%\fmfiv{d.sh=cross,d.ang=0}{point length(p1)/3 of p1}
\end{fmfgraph*}
\end{fmffile}
\end{tabular}
\end{center}
\caption{Two schematic amplitudes whose dimension $n$ parts are equal: the square and circle denote higher and lower derivative dimension $n$ operators. \label{fig:eomgraph}}
\end{figure}

The dimension $n > 4$ part of the following two tree level graphs are equivalent, when all external legs are on-shell: a) a graph comprising a dimension $n$ vertex and a dimension 4 vertex, and b) a graph comprising a single dimension $n$ vertex with the same external legs. This is illustrated schematically in Figure \ref{fig:eomgraph}. At the operator level, this corresponds to redundancies amongst dimension $n$ operators arising from our freedom to make field redefinitions \cite{Arzt:1993gz}. The redundancies take the form
\begin{equation}
U(x) \frac{\del S_4}{\del F(x)} = 0,
\label{eq:eomrel}
\end{equation}
where $U(x)$ is a functional of some fields, which depend on spacetime coordinate $x$, and $\frac{\del S_4}{\del F(x)}$ is an equation of motion (EOM) of the renormalizable theory: the functional derivative of the dimension 4 action w.r.t. a constituent field $F$. Note that $\dim U + 4 - \dim F = n$. Higher dimension ($> n$) components of (\ref{eq:eomrel}) have been neglected.

\DEFT\ constructs the EOM according to the following functional derivative rules:
\begin{equation}
\frac{\del (D D \ldots D) H(x)}{\del H(y)} = (D D \ldots D) \del^{(4)}(x-y) 
\end{equation}
for any field $H$;
\begin{equation}
\frac{\del D_{\al \dot \al} H^a}{\del {A_{\be \dot \be}}^c_d(x)} = i g_N \del^\be_\al \del^{\dot \be}_{\dot \al} \left( \del^a_c \del^d_b H^b - \frac{1}{N} \del^d_c H^a \right) \del^{(4)}(x-y)
\end{equation}
for a field $H^a$ charged under an $SU(N)$ gauge group, with coupling $g_N$ and vector potential $A^c_d$;
\begin{equation}
\frac{\del {F_{\al \be}}^a_b}{\del {A_{\ga \dot \ga}}^c_d(x)} = \eps^{\dot\al \dot\be} \del^a_c \del^d_b \left( \del^\ga_\be \del^{\dot \ga}_{\dot \be} D_{\al \dot\al}- \del^\ga_\al \del^{\dot \ga}_{\dot \al} D_{\be \dot\be} \right) \del^{(4)}(x-y) 
\end{equation}
for the field strength $F^a_b$ of the $SU(N)$ gauge group, vector potential $A^c_d$.

For each monomial element of an EOM, the program searches for its embedding in each dimension $n$ term, calculating corresponding `quotients' $U(x)$ (\ref{eq:eomrel}). Then, for each EOM and each possible $U(x)$, relations are formed out of the corresponding terms, weighted by the coefficients of the EOM.

%%%%%%%%%%%%%%%%%%%%%%%%%%%%%%%%%%%%%%%%%%%%%%%%%
\subsection{Using the code \label{sec:user}}
%%%%%%%%%%%%%%%%%%%%%%%%%%%%%%%%%%%%%%%%%%%%%%%%%%

\newcommand{\term}{\texttt{Term}~}
\newcommand{\terms}{\texttt{Term}s~}
\newcommand{\field}{\texttt{Field}~}
\newcommand{\invariant}{\texttt{Invariant}~}
\newcommand{\invariants}{\texttt{Invariant}s~}
\newcommand{\relation}{\texttt{Relation}~}
\newcommand{\ind}{\texttt{Index}~}
\newcommand{\frc}{\texttt{frac}~}
\newcommand{\symmetry}{\texttt{Symmetry}~}
\newcommand{\sympy}{\texttt{sympy}~}

The code requires \texttt{Python 2.7+} and the \texttt{sympy} \cite{10.7717/peerj-cs.103} symbolic manipulation package.

Each monomial operator is represented by a \term instance. A \term has a list (\texttt{.\_fields}) of \field instances and a list (\texttt{.\_invariants}) of \invariant instances, each of which has a list (\texttt{.indices}) of \ind instances that they respectively possess or contract. \field instances also have a list (\texttt{.Dindices}) of tuples of its indices which belong to covariant derivatives acting on the field; a dictionary (\texttt{.U1Dict}) of $U(1)$ charges, which take on rational values represented by \frc instances, and a list (\texttt{.symmetries}) of \symmetry instances which enforce symmetry properties of the field's indices upon contraction with \invariants.

A \relation is a list of \terms (\texttt{.terms}) and corresponding coefficients (\texttt{.weights}), which are \sympy expressions. \sympy is used for some of the subsequent matrix manipulation.\footnote{The row reduction, when performed symbolically with the marginal couplings of the theory as variables, is computationally expensive. One has the option in \DEFT\ of substituting the different marginal couplings for prime numbers to speed up the row operations, or substituting for zeroes (i.e.~working with a free renormalizable part of the theory). Note that one should avoid replacing the couplings with floating point values prior to the row reduction, due to the ensuing propagation of floating point inaccuracies.}

The use of the provided methods for the generation of terms and relations, as well as the conversion into and between bases, is documented in the unit tests, which compute the cross checks described in \S\ref{sec:cross}.

%%%%%%%%%%%%%%%%%%%%%%%%%%%%%%%%%%%%%%%%%%%%%%%%%%
\section{Cross checks \label{sec:cross}}
%%%%%%%%%%%%%%%%%%%%%%%%%%%%%%%%%%%%%%%%%%%%%%%%%%
%%%%%%%%%%%%%%%%%%%%%%%%%%%%%%%%%%%%%%%%%%%%%%%%%%
\subsection{Dimensions \label{sec:dim}}
%%%%%%%%%%%%%%%%%%%%%%%%%%%%%%%%%%%%%%%%%%%%%%%%%%
\begin{figure}
\begin{center}
\begin{tabular}{ccccccccc}
$d$ &1&2&3&4&5&6&7&8\\ \hline
$\{H,l_L\}$ &0&1&0&3&2&6&6&18 \\
$\{\phi,H,l_L\}$&1&2&2&6&5&12&21&48 \\
$\{B,e_R\}$&0&0&0&3&0&1&0&5 \\
$\{H,B,W,l_L,e_R\}$&0&1&0&10&2&23&12&179 \\
$\{H,B,W,G,l_L,e_R,q_L,u_R,d_R\}$&0&1&0&19&2&84&30&993 \\
$\{H,B,W,l_L,e_R,q_L^4,u_R^4,d_R^4\}$&0&1&0&17&2&68&*&* \\
$\{H,B,W,G^4,l_L,e_R,q_L^4,u_R^4,d_R^4\}$&0&1&0&19&2&76&*&* 
\end{tabular}
\end{center}
\caption{The number of independent operators at each mass dimension
  $d$, for various combinations of
  fields. $\{H,B,W,G,l_L,e_R,q_L,u_R,d_R\}$ are those of the one
  generation Standard Model (cf. Figure \ref{fig:smfieldreps}); $\phi$
  is a real scalar singlet under the symmetries of the Standard
  Model. $G^4$ and $\{q_L^4,u_R^4,d_R^4\}$ are respectively the gauge
  boson and matter fields of an $SU(4)$ gauge group, with the same
  electroweak charges as their $SU(3)$ charged counterparts in the
  Standard Model. In the penultimate line of the Table, we treat
    the $SU(4)$ as a global symmetry.\label{tab:crosschecks}}
\end{figure}

We used the code to calculate the number of independent operators at
each mass dimension up to 8, for lagrangians containing various
combinations of light fields. The results agree with Figure
\ref{tab:crosschecks}, whose entries were either computed manually ($d
\leq 4$) or using the Hilbert series method of \cite{Henning:2015alf} ($d >
4$).\footnote{Both \DEFT\ and Table \ref{tab:crosschecks} count the
  dimension 4 operators $F_{\al \be} F^{\al \be}$ and $\bar F_{\dot\al
    \dot\be} \bar F^{\dot\al \dot\be}$ independently for any field
  strength $F$.}
%%%%%%%%%%%%%%%%%%%%%%%%%%%%%%%%%%%%%%%%%%%%%%%%%%
\subsection{Change of basis\label{sec:cob}}
%%%%%%%%%%%%%%%%%%%%%%%%%%%%%%%%%%%%%%%%%%%%%%%%%%

We define the one generation SILH basis as a one generational
restriction of the operators in Tables 1, 2, and 3 of
\cite{Falkowski:2001958},\footnote{The operators $O_{Hl}$,
  $O_{Hl}^\prime$, $O_{ll}$, $O_{lu}$, and $O_{uu}$ are absent.} and
similarly the one generation Warsaw basis from Tables 2 and 3 of
\cite{Grzadkowski:2010es}. \DEFT\ generates expressions for the Wilson coefficients of the SILH basis in terms of the equivalent Wilson coefficients of the Warsaw basis, in agreement with the one generational restriction of the formulae of Appendix A of \cite{Falkowski:2001958}, as well as an independent manual calculation.

%%%%%%%%%%%%%%%%%%%%%%%%%%%%%%%%%%%%%%%%%%%%%%%%%%
\section{Discussion \label{sec:disc}}
%%%%%%%%%%%%%%%%%%%%%%%%%%%%%%%%%%%%%%%%%%%%%%%%%%
In an auxiliary directory in the submission, we provide a list of possible monomial operators in the one generation Standard Model, in the default ordering,\footnote{The operators are listed in descending order of number of derivatives, then number of epsilon tensors, indices and conjugate fields.} together with the reduced row echelon form of $M_{ij}$ (the equivalent of the RHS of (\ref{eq:rrefexample})) when the columns are so ordered.

\begin{figure}
\includegraphics[width=\textwidth]{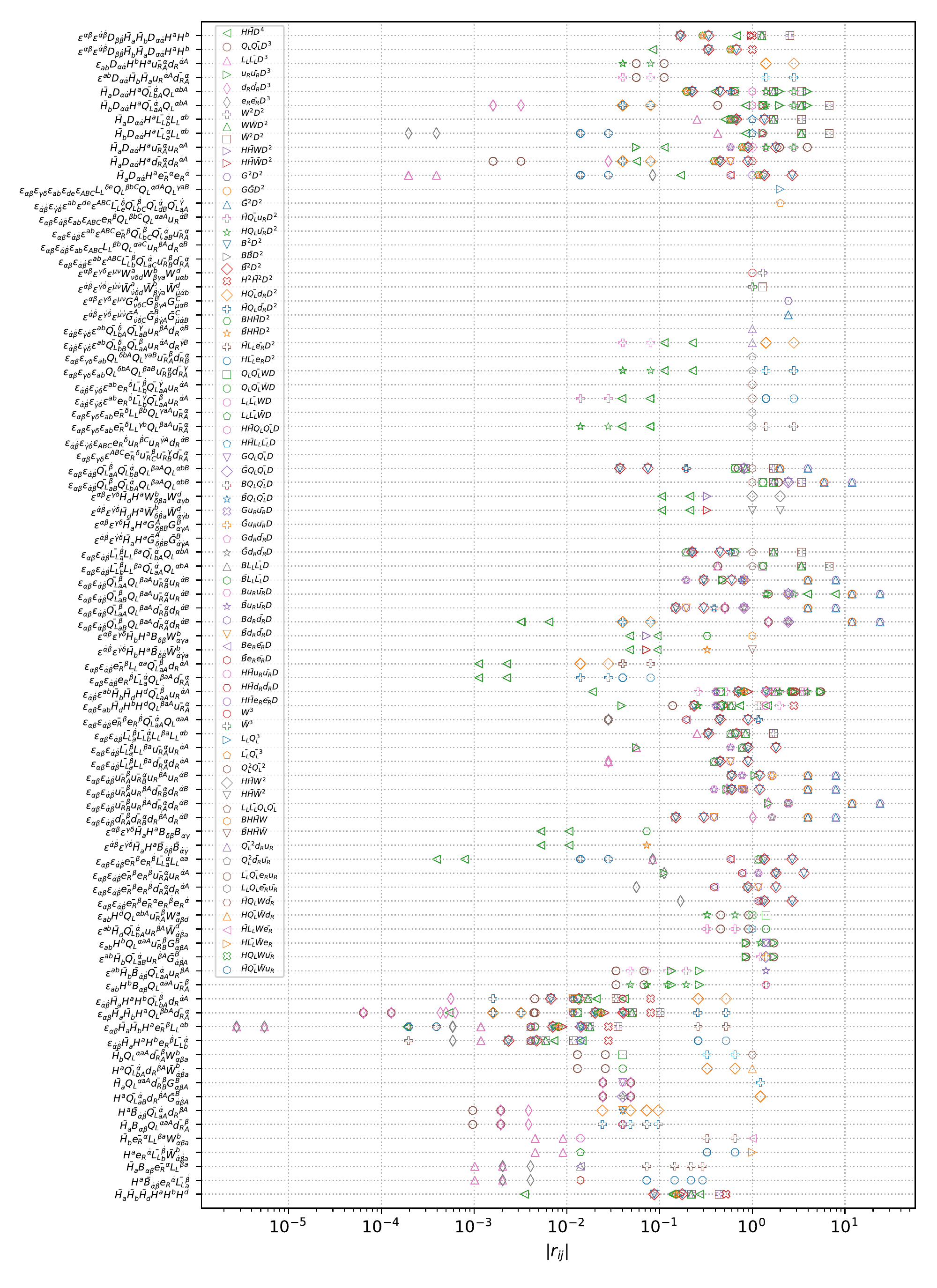}
\caption{The absolute values of the non-trivial components of $r_{ij}$, defined in (\ref{eq:rrefexample}), for the one generation SM. Each marker is positioned in line with the basis operator corresponding to column $j$, and formatted according to the field composition of the redundant operator corresponding to the leading coefficient in the row $i$. We define the marginal couplings in terms of the measured coefficients of the three generation Standard Model: the Higgs quartic and gauge couplings equal those measured at the $Z$ pole, whereas the Yukawa couplings are set by measurements of the heaviest generation: $y_u = \frac{m_t}{v}$, $y_d = \frac{m_b}{v}$, $y_e = \frac{m_\tau}{v}$. \label{fig:crappamundi}}
\end{figure}

We depict the structure of the RREF matrix in Figure \ref{fig:crappamundi}, by plotting the non-trivial values of $\abs{r_{ij}}$, as defined in (\ref{eq:rrefexample}). Each row of the RREF matrix effectively defines a linear relation expressing each redundant operator in terms of the remaining basis operators; therefore, for each row $i$, we plot the absolute values $\abs{r_{ij}}$ on the line of the basis operator corresponding to the index $j$. For each row, the style of marker is determined by the field composition of the redundant operator `being eliminated', as indicated in the legend. Note that we have chosen to relax the hermiticity condition on the operators, such that many will have, in general, complex Wilson coefficients. The components $r_{ij}$ are in general complex.

Note that there are eight lines in Figure \ref{fig:crappamundi} with no points; equivalently, there are eight columns in the RREF matrix whose entries are all zero, corresponding to six $B$ violating operators and the two operators of the form $H \bar H G^2$ and $H \bar H \bar G^2$. These eight monomial operators are the sole monomial representatives of their respective equivalence classes, and are, therefore, always in any basis constructed solely from monomial operators. In addition, the remaining two $B$ violating operators --- $Q^3 L$ and ${\bar Q}^3 \bar L$ --- are only somewhat trivially related to operators of the same field composition via a Fierz relation. By considering the structure of the redundancy relations, we provide some justification for the apparent isolation of these ten operators in Appendix \ref{app:relstruct}.

\acknowledgments
We thank T.~ You and other members of the Cambridge SUSY Working group for discussions. 
BG was supported by STFC grants ST/L000385/1 and ST/P000681/1 and King's College,
Cambridge. DS acknowledges support from the Science and Technology Facilities Council; Emmanuel College, Cambridge; the Department of Energy (DE-SC0014129), and the Center for Scientific Computing from the CNSI, MRL: an NSF MRSEC (DMR-1121053) and NSF CNS-0960316.

\appendix
\section{Index conventions\label{app:conv}}

For an $N$ index group
\begin{equation}
\eps^{12 \cdots N} = - \eps_{12 \cdots N} = 1; \qquad \del^a_b = \begin{cases} 1 & \text{if $a=b$,} \\ 0 & \text{otherwise.} \end{cases}
\end{equation}
We use the spinor index conventions of \cite{Dreiner:2008tw} with a mostly-minus metric $\eta_{\mu\nu}=\mathrm{diag}(+1,-1,-1,-1)$ and totally antisymmetric tensor $\eps^{0123}=-\eps_{0123}=+1$. With the use of the tensors $\sigma^\mu_{\al\dot\al}$ and $\bar \sigma_\mu^{\dot\al\al}$ and relations (2.47-2.53) of \cite{Dreiner:2008tw}
\newcommand{\sigmabar}{\bar \sigma}
\newcommand{\metric}{\eta}
\begin{align}
\sigma^\mu_{\alpha\dot{\alpha}} \sigmabar_\mu^{\dot{\beta}\beta}
= 2
\delta_{\alpha}{}^{\beta} \delta^{\dot{\beta}}{}_{\dot{\alpha}}\,,
\\
\sigma^\mu_{\alpha\dot{\alpha}} \sigma_{\mu\beta\dot{\beta}}
= 2
\epsilon_{\alpha\beta} \epsilon_{\dot{\alpha}\dot{\beta}}\,,
\\
\sigmabar^{\mu\dot{\alpha}\alpha} \sigmabar_\mu^{\dot{\beta}\beta}
= 2
\epsilon^{\alpha\beta} \epsilon^{\dot{\alpha}\dot{\beta}}\,,
\\
{[\sigma^\mu\sigmabar^\nu + \sigma^\nu \sigmabar^\mu ]_\alpha}^\beta
= 2\metric^{\mu\nu} \delta_{\alpha}{}^{\beta}\,,
\\
[\sigmabar^\mu\sigma^\nu + \sigmabar^\nu \sigma^\mu
]^{\dot{\alpha}}{}_{\dot{\beta}}
= 2\metric^{\mu\nu} \delta^{\dot{\alpha}}{}_{\dot{\beta}}\,,
\\
\sigma^\mu \sigmabar^\nu \sigma^\rho =
\metric^{\mu\nu} \sigma^\rho
-\metric^{\mu\rho} \sigma^\nu
+\metric^{\nu\rho} \sigma^\mu
+i \epsilon^{\mu\nu\rho\kappa} \sigma_\kappa\,,
\\
\sigmabar^\mu \sigma^\nu \sigmabar^\rho =
\metric^{\mu\nu} \sigmabar^\rho
-\metric^{\mu\rho} \sigmabar^\nu
\metric^{\nu\rho} \sigmabar^\mu
-i \epsilon^{\mu\nu\rho\kappa} \sigmabar_\kappa\,,
\end{align}
expressions involving vector and spinor Lorentz indices may be easily converted. For expressions with a single vector index we define
\begin{equation}
A_{\al\dot\al} = \sigma^\mu_{\al\dot\al} A_\mu , \qquad 
D_{\al\dot\al} = \sigma^\mu_{\al\dot\al} D_\mu ,
\end{equation}
whence we derive
\begin{equation}
A_\mu = \frac{1}{2} \bar\sigma_{\mu}^{\dot\al \al} A_{\al \dot \al} , \qquad
D_\mu = \frac{1}{2} \bar\sigma_{\mu}^{\dot\al \al} D_{\al \dot \al} .
\end{equation}
For expressions with two vector indices, such as a field strength $F_{\mu\nu}$ or its dual $\tilde F_{\mu\nu}$, we define
\begin{equation}
F_{\al\dot\al\be\dot\be} = \sigma^\mu_{\al\dot\al} \sigma^\nu_{\be\dot\be} F_{\mu\nu} , \qquad
\tilde F_{\al\dot\al\be\dot\be} = \sigma^\mu_{\al\dot\al} \sigma^\nu_{\be\dot\be} \tilde F_{\mu\nu} ,
\end{equation}
such that
\begin{equation}
F_{\mu\nu} = \frac{1}{4} \bar \sigma_\mu^{\dot\al\al} \bar \sigma_\nu^{\dot\be\be} F_{\al\dot\al\be\dot\be} , \qquad
\tilde F_{\mu\nu} = \frac{1}{4} \bar \sigma_\mu^{\dot\al\al} \bar \sigma_\nu^{\dot\be\be} \tilde F_{\al\dot\al\be\dot\be} \, ,
\end{equation}
where $F_{\al\dot\al\be\dot\be}$ and $\tilde F_{\al\dot\al\be\dot\be}$ may be expressed in terms of Lorentz group irreps $F_{\al\be}$ and $\bar F_{\dot\al\dot\be}$:
\begin{equation}
F_{\al\dot\al\be\dot\be} = -\frac{1}{2} (\eps_{\dot\al\dot\be} F_{\al\be} + \eps_{\al\be} \bar F_{\dot\al\dot\be}) , \qquad
\tilde F_{\al\dot\al\be\dot\be} = -\frac{1}{2} i (\eps_{\dot\al\dot\be} F_{\al\be} - \eps_{\al\be} \bar F_{\dot\al\dot\be}).
\end{equation}
If $F_{\mu\nu} = \partial_\mu A_\nu - \partial_\nu A_\mu$, helpful consequences of the above conventions include
\begin{align}
F_{\al\be} = \eps^{\dot\al\dot\be} F_{\al\dot\al\be\dot\be} = \eps^{\dot\al\dot\be} (\partial_{\al\dot\al} A_{\be\dot\be} - \partial_{\be\dot\be} A_{\al\dot\al}), \\
\bar F_{\dot\al\dot\be} = \eps^{\al\be} F_{\al\dot\al\be\dot\be} = \eps^{\al\be} (\partial_{\al\dot\al} A_{\be\dot\be} - \partial_{\be\dot\be} A_{\al\dot\al}) .
\end{align}

Note that, alternatively, one can use the tensors
\begin{align}
(\sigma^{\mu\nu})_\al^\be = \frac{1}{4} i (\sigma^\mu_{\al\dot\al} \bar \sigma^{\nu \, \dot\al \be} - \sigma^\nu_{\al\dot\al} \bar \sigma^{\mu \, \dot\al \be}) \\
(\bar \sigma^{\mu\nu})^{\dot\al}_{\dot\be} = \frac{1}{4} i (\bar \sigma^{\mu \, \dot\al\al} \sigma^{\nu}_{\al \dot\be} - \bar \sigma^{\nu \, \dot\al\al} \sigma^{\mu}_{\al \dot\be})
\end{align}
to convert directly between different forms of the field strength:
\begin{align}
F_{\al\be} = 2 i (\sigma^{\mu\nu})_\al^\ga \eps_{\ga\be} F_{\mu\nu}; \qquad \bar F_{\dot\al\dot\be} = 2 i \eps_{\dot\al\dot\ga} (\bar \sigma^{\mu\nu})^{\dot\ga}_{\dot\be} F_{\mu\nu}; \\
F^{\mu\nu} - i \tilde F^{\mu\nu} = -\frac{1}{2} i F_{\al\ga} \eps^{\ga\be} (\sigma^{\mu\nu})_\beta^\al; \qquad
F^{\mu\nu} + i \tilde F^{\mu\nu} = -\frac{1}{2} i \eps^{\dot\al\dot\ga} \bar F_{\dot\ga\dot\be} (\bar \sigma^{\mu\nu})^{\dot\beta}_{\dot\al} \, .
\end{align}

A four component Dirac spinor may be expanded in terms of the components of a left-handed, $L^\al$, and right-handed, $R^{\dot\al}$, Weyl spinor, such that
\begin{equation}
\Psi = \begin{pmatrix} L_\al \\ R^{\dot\al} \end{pmatrix} = {\scriptsize \begin{pmatrix} L_1 \\ L_2 \\ R^1 \\ R^2 \end{pmatrix} },
\end{equation}
and its conjugates are
\begin{equation}
\bar \Psi = \begin{pmatrix} \bar R^\al & \bar L_{\dot\al} \end{pmatrix}; \qquad \Psi^C = \begin{pmatrix} \bar R_\al \\ \bar L^{\dot\al} \end{pmatrix} \, . 
\end{equation}
Gamma matrices may be similarly expanded as
\begin{equation}
\gamma^\mu = \begin{pmatrix} 0 & \sigma^\mu_{\al \dot \be} \\ \bar \sigma^{\mu \dot \al \be} & 0 \end{pmatrix}; \qquad
\gamma^5 = \begin{pmatrix} -\del_\al^\be & 0 \\ 0 & \del^{\dot\al}_{\dot\be} \end{pmatrix}; \qquad
\frac{1}{4} i [\gamma^\mu,\gamma^\nu] = \begin{pmatrix} (\sigma^{\mu\nu})_\al^\be & 0 \\ 0 & (\bar \sigma^{\mu\nu})^{\dot\al}_{\dot\be} \end{pmatrix} \, .
\end{equation}

We normalize the non-Abelian vector potentials of the SM such that
\begin{equation}
(W_{\al\dot\al})^a_b = \frac{1}{2} W_{\al\dot\al}^i (\sigma^i)^a_b; \qquad
(G_{\al\dot\al})^a_b = \frac{1}{2} G_{\al\dot\al}^i (\lambda^i)^a_b \, ,
\end{equation}
where $(\lambda^i)^a_b$ is the value of the $a$th row and $b$th column of the $i$th Gellmann matrix, and similarly for the Pauli sigma matrices $(\sigma^i)^a_b$. $G^i$, $i=1,\ldots,8$, and $W^i$, $i=1,\ldots,3$, are the canonical gauge fields found in, for instance, the listing of the Warsaw basis \cite{Grzadkowski:2010es}. With the use of the Fierz relations,
\begin{align}
(\sigma^i)^a_b (\sigma^i)^c_d &= 2 \delta^a_d \delta^c_b - \delta^a_b \delta^c_d \\
(\lambda^i)^a_b (\lambda^i)^c_d &= 2 \delta^a_d \delta^c_b - \frac{2}{3} \delta^a_b \delta^c_d
\end{align}
we can deduce the correct normalization of the kinetic terms, e.g.,
\begin{align}
-\frac{1}{4} W^i_{\mu\nu} W^{i \, \mu\nu} = \frac{1}{16} \left( (W^{\al\be})^a_b (W_{\al\be})^b_a + (\bar W^{\dot\al\dot\be})^a_b (\bar W_{\dot\al\dot\be})^b_a \right) \, ,\\
-\frac{1}{4} G^i_{\mu\nu} G^{i \, \mu\nu} = \frac{1}{16} \left( (G^{\al\be})^a_b (G_{\al\be})^b_a + (\bar G^{\dot\al\dot\be})^a_b (\bar G_{\dot\al\dot\be})^b_a \right) \, .
\end{align}

\section{The structure of operator relations in a generic 4d EFT \label{app:relstructure}}
\label{app:relstruct}

\begin{figure}
\def\eps{0.32}
\def\eoms{->,blue}
\begin{tikzpicture}[x=3.5cm,y=2cm]
\node (d4s2) at (2,0-\eps) {$\phi^2 D^4$};
\node (d3ppb) at (2,0+\eps) {$\psi \bar \psi D^3$};
\node (d2ffb) at (2,0) {$F \bar F D^2$};
\node (d2s4) at (4,0-\eps) {$\phi^4 D^2$};
\node (p2pb2) at (4,0+\eps) {$\psi^2 {\bar \psi}^2$};
\node (dppbs2) at (4,0) {$\psi \bar \psi \phi^2 D$};
\node (s6) at (6,0) {$\phi^6$};
\node (d2f2) at (2,2) {$F^2 D^2$};
\node (f2s2) at (4,2-\eps) {$F^2 \phi^2$};
\node (fp2s) at (4,2+\eps) {$F \psi^2 \phi$};
\node (p4) at (4,2) {$\psi^4$};
\node (dfppb) at (3,1+\eps) {$F \psi \bar \psi D$};
\node (d2p2s) at (3,1) {$\psi^2 \phi D^2$};
\node (d2fs2) at (3,1-\eps) {$F \phi^2 D^2$};
\node (f3) at (3,3) {$F^3$};
\node (p2s3) at (5,1) {$\psi^2 \phi^3$};
\foreach \start/\end in {d4s2/d2s4,d2s4/s6,d2p2s/p2s3,d4s2/d2p2s,d2p2s/p4,d2p2s/p2pb2}
{
\draw[->,blue,thick] (\start.east) -- (\end.west);
}
\foreach \start/\end in {d3ppb/d2p2s,dfppb/fp2s,d2p2s/dppbs2,dppbs2/p2s3}
{
\draw[->,green,dashed,thick] (\start.east) -- (\end.west);
}
\foreach \start/\end in {d2f2/dfppb,d2f2/d2fs2,d2ffb/dfppb,d2ffb/d2fs2,dfppb/p2pb2,dfppb/dppbs2,d2fs2/dppbs2,d2fs2/d2s4}
{
\draw[->,red,dash pattern=on 8pt off 2pt,thick] (\start.east) -- (\end.west);
}
\foreach \start/\end in {d2f2/f3,d4s2/d2fs2,d3ppb/dfppb,d2p2s/fp2s,d2fs2/f2s2}
{
\draw[->,gray,dashdotted,thick] (\start.east) -- (\end.west);
}
\draw[->,thick] (1.6,-1) -- (6.2,-1) node [right] {$n$};
\draw[->,thick] (1.6,-0.5) -- (1.6,3.5) node [above] {$\sum h$};
\foreach \x in {2,3,4,5,6} {
\draw (\x,-0.8) -- (\x,-1.2) node [below] {\x};
}
\foreach \y in {0,1,2,3} {
\draw (1.7,\y) -- (1.5,\y) node [left] {\y};
}
\end{tikzpicture}
\caption{A schematic `map' of all dimension 6 operators allowed by Lorentz symmetry, cut in half about its axis of symmetry $\sum h =0$ (reflected in this line are the hermitian conjugates of the operators shown). Arrows show the movement induced by equation of motion relations and commutation of derivative relations in the space of dim 6 operators, colour-coded to show \textcolor{blue}{$\phi$ EOMs (blue)}, \textcolor{green}{$\psi$ EOMs (green, short dashed)}, \textcolor{red}{$F$ EOMs (red, long dashed)}, and \textcolor{gray}{replacing derivatives with field strengths (grey, dash-dotted)}.
\label{fig:moves}}
\end{figure}
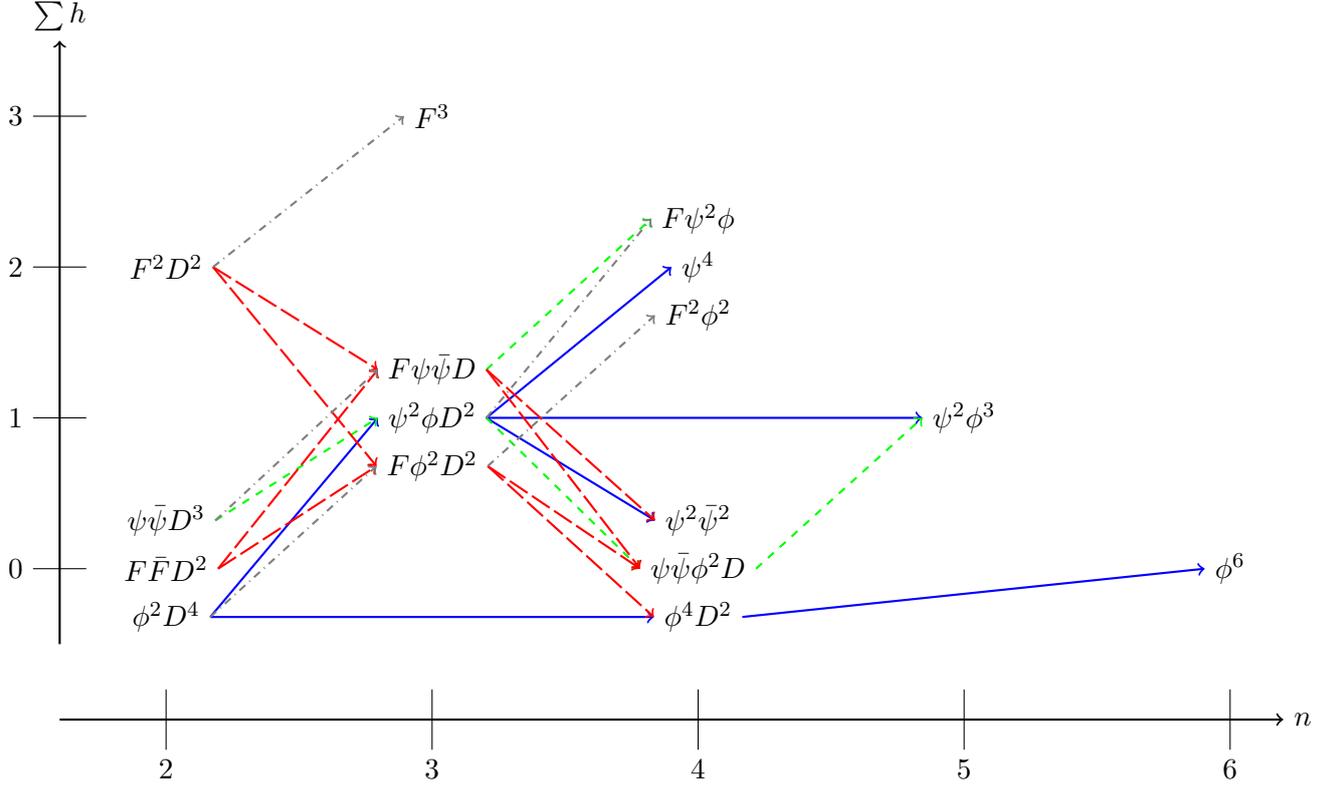

Following the procedure of \cite{Cheung:2015aba}, we define two integer coordinates $n$ and $\sum h$ for each monomial EFT operator as, respectively, the number of fields and the sum of the helicities $h$ of the particle created by the action of each field on the vacuum.\footnote{For the purposes of calculating $n$ and $\sum h$, we treat covariant derivatives as partial derivatives.} For fields that are scalar $\phi$, left- and right-handed Weyl fermions $\psi$ and $\bar \psi$, and left- and right-handed field strengths $F$ and $\bar F$, $h=0,+\frac{1}{2},-\frac{1}{2},+1,-1$ respectively. We enumerate the possible field compositions of dimension 6 operators allowed by Lorentz symmetry and arrange them by their coordinates in Fig.~\ref{fig:moves} (cf.~Fig.~1 of \cite{Cheung:2015aba}).

Consider how redundancy relations allow one to move around the table of Fig.~\ref{fig:moves}. IBP and Fierz relations `trivially' mix operators with the same field composition, and therefore with the same coordinates $(n,\sum h)$. The other two kinds can be viewed as expressing a higher derivative operator in terms of an equivalent sum of lower derivative ones.

One, replacing a commutator of derivatives with a field strength generically yields a combination of terms, some with an additional $F$, some with an $\bar F$ (one of these may be forbidden by Lorentz symmetry). Thus, starting with an operator with coordinates $(n,\sum h)$, one ends up with operators at $(n+1,\sum h + 1)$ and $(n+1,\sum h -1)$.

Two, replacing the free part of an EOM with the interacting parts amounts to, diagrammatically, taking a graph comprising just an insertion of a higher derivative dim 6 operator, and adding a dim 4 vertex to one of the legs on which the derivative(s) act(s) (see Figure \ref{fig:eomgraph}). This composite, two vertex graph may have the same leading order momentum piece as a simple insertion of a lower derivative dim 6 operator, when the external legs are on shell. We may assume the fields are massless, as relevant interactions do not affect the EOM relations.\footnote{More precisely, the effects of mass terms in the EOMs can be absorbed by redefinitions of the dim 4 coefficients in the lagrangian.} Thus, by (12) of \cite{Cheung:2015aba}, the coordinates $(n_i,(\sum h)_i)$ of such a composite amplitude (and by extension the weights of the corresponding lower derivative operator) are related to the weights of the constituent vertices $(n_j,(\sum h)_j)$ and $(n_k,(\sum h)_k)$ by:
\begin{equation}
n_i = n_j + n_k -2 ; \qquad
(\sum h)_i = (\sum h)_j + (\sum h)_k .
\end{equation}

The weights of possible dim 4 vertices are as follows. A gauge or Yukawa coupling is $(n,\sum h)=(3,\pm 1)$. Anything proportional to a scalar quartic is $(4,0)$. Therefore, the part of an EOM relation proportional to a gauge or Yukawa coupling lies one unit right and one unit either up or down in the table of operators, relative to the original higher derivative operator. For the part proportional to a Higgs quartic, it lies two units to the right.

Figure \ref{fig:moves} allows us to understand two examples of dimension 6 monomial operators in the SM, which are not related to any others. One, an $H^2 G^2$ (class $F^2 \phi^2$) operator could only be reached from an operator of class $F \phi D^2$. However, all such $G H D^2$ operators are forbidden by gauge symmetries. Two, baryon violating operators of the form $\psi^4$, $\psi^2 \bar \psi^2$, and $\bar \psi^4$, are only reachable from operators of class $F \psi \bar \psi D$ and $\psi^2 D^2$, as well as their conjugates. The baryon violating operators contain three quarks, and all relations preserve the parity of the number of quarks. However, there are no gauge invariant operators of the form $F \psi \bar \psi D$ or $\psi^2 D^2$ containing a single quark, leaving the baryon violating four fermion operators unrelated to both baryon conserving operators, and also unrelated to each other.

\bibliographystyle{JHEP}
\bibliography{operators}

\end{document}